\documentclass[12pt]{article}
\usepackage{amsmath} 
\usepackage[utf8]{inputenc} 
\usepackage{url} 
\usepackage{booktabs} 
\usepackage{caption} 
\usepackage{geometry} 
\usepackage{hyperref} 
\usepackage{graphicx}
\usepackage{subcaption}
\usepackage{float}
\usepackage{textcomp} 
\title{Private MEV Protection RPCs: Benchmark Study}
\date{24 February 2025}

\begin{document}

\begin{center}
  \vspace*{0.5cm} 
  \Large \textbf{Private MEV Protection RPCs: Benchmark Study} \\[0.3cm] 
  \normalsize 24 February 2025 
\end{center}

\vspace{0.3cm} 
\begin{center}
  \begin{minipage}{0.45\textwidth}
    \centering
    {Alex Vinyas} \\[0.2cm]
    CoW DAO \\[0.2cm]
    \texttt{alex.vinyas@cow.fi}
  \end{minipage}
  \hfill
  \begin{minipage}{0.45\textwidth}
    \centering
    {Paul Janicot} \\[0.2cm]
    CoW DAO \\[0.2cm]
    \texttt{paul@cow.fi}
  \end{minipage}
\end{center}
\vspace{0.5cm} 

\begin{abstract}
Decentralized Finance (DeFi) on Ethereum has undergone significant transformations since its emergence during the DeFi summer of 2020. With the introduction of Proof of Stake (PoS) and Proposer-Builder Separation (PBS), the transaction supply chain on Ethereum has shifted from relying entirely on the public mempool for DeFi interactions to an astonishing 80\% usage of private RPCs~\cite{private_mempool_query}. These private RPCs submit transactions directly to builders therefore skipping the public mempool, while  conducting Order Flow Auctions (OFAs) to capture MEV backrun rebates and gas rebates. However, not all private RPCs  employ the same mechanism  or achieve similar outcomes, despite each claiming to be ''the best''.
In this study, we conduct an empirical evaluation of the quality and effectiveness of four major OFAs within Ethereum— MEV Blocker, Flashbots Protect, Blink, and Merkle—by simultaneously submitting identical transactions to each RPC and measuring the results. Our findings reveal that not all RPCs OFAs produce the same outcomes. These insights underscore the significant implications of OFA design choices on transaction efficiency and execution quality, and thus why an order flow originators should pay close attention to which OFA they use.

\end{abstract}

\section{Introduction}
The Ethereum public mempool serves as a crucial component of its decentralized transaction processing system, acting as a waiting area where pending transactions are broadcast before being included in a block. However, this open and transparent structure introduces several vulnerabilities that have led to the emergence of sophisticated transaction exploitation strategies, collectively known as Maximal Extractable Value (MEV) attacks. These forms of MEV extraction not only degrade user experience but also increase transaction costs, contribute to network congestion, and compromise the fairness of DeFi markets.
To mitigate these challenges, a growing number of MEV protection solutions have emerged, most notably private mempools implemented via RPCs. These RPCs offer alternative transaction submission mechanisms that bypass the public mempool, preventing actors from exploiting transaction visibility. Furthermore, by leveraging Order Flow Auctions (OFAs), MEV protection RPCs aim to redistribute MEV more equitably, offering users improved transaction execution and potential rebates from backrun profits as well as gas optimizations. However, despite their shared goal of mitigating MEV risks, different MEV protection RPCs vary significantly in their mechanisms, effectiveness, and economic incentives.
We perform an empirical comparison of four major MEV protection RPCs—MEV Blocker, Flashbots, Blink, and Merkle—by evaluating their execution quality for the exact same transactions. Through a systematic analysis of transaction outcomes, we examine how their different OFA designs impact user benefits, shedding light on how order flow providers should carefully pay attention to marketing claims and thoroughly review to whom they are sending their transactions to.

\subsection{Private RPCs as a Solution}
One of the most effective ways to mitigate front-running and other forms of MEV exploitation is through private mempools offered by MEV Protection RPCs. These systems allow users to submit transactions directly to block builders or route them through specialized Order Flow Auctions (OFAs), thereby bypassing the public mempool where transactions are vulnerable to exploitation. By keeping transactions private until execution, MEV Protection RPCs prevent front-running and ensure that users can leverage searchers to their advantage, rather than having them extract value from them.

\begin{figure}[H]
    \centering
    \includegraphics[width=0.85\textwidth]{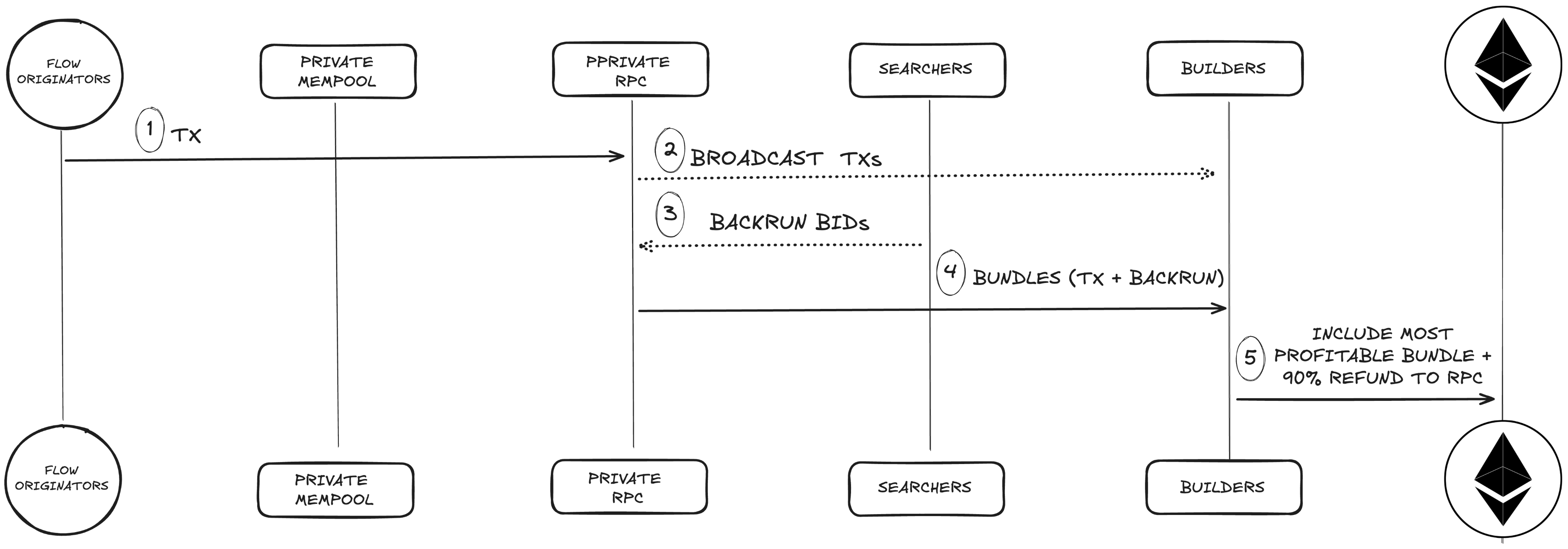}
    \caption{Transaction flow in MEV protection RPCs.}
    \label{fig:ofa_flow}
\end{figure}

A key component of this system is the role searchers play, as MEV Protection RPCs reorient their incentives, encouraging them to facilitate rebates for users. Within an OFA framework, searchers compete to improve trade outcomes for users by eliminating the predatory behavior associated with front-running and redistributing a portion of the MEV back to the original traders in the form of rebates.

In practice, MEV Protection RPCs integrate searchers into their transaction processing pipelines in a way that benefits unsophisticated traders, who may lack the resources or expertise to optimize their own execution strategies. Instead of submitting transactions to the public mempool, users can leverage these private RPCs to access searchers who compete within an auction environment to construct backrun bundles. A user’s transaction is paired with a backrunning transaction that extracts MEV, and the searcher submits a bid representing the portion of that value they are willing to share in exchange for inclusion. These bundles are then relayed to block builders, who prioritize the most profitable bids when constructing blocks.

Once a winning bundle is selected, the bid amount is split—commonly allocating a majority (e.g., 90\%) to the user and a smaller portion (e.g., 10\%) to the builder.

\section{Main Providers and Differences Among OFAs}
Within the context of MEV Protection RPCs and their respective OFAs, we evaluate the differences in mechanisms, incentives, and execution flow among four providers: Flashbots, Merkle, Blink, and MEV Blocker.

\begin{figure}[H]
    \centering
    \includegraphics[width=0.85\textwidth]{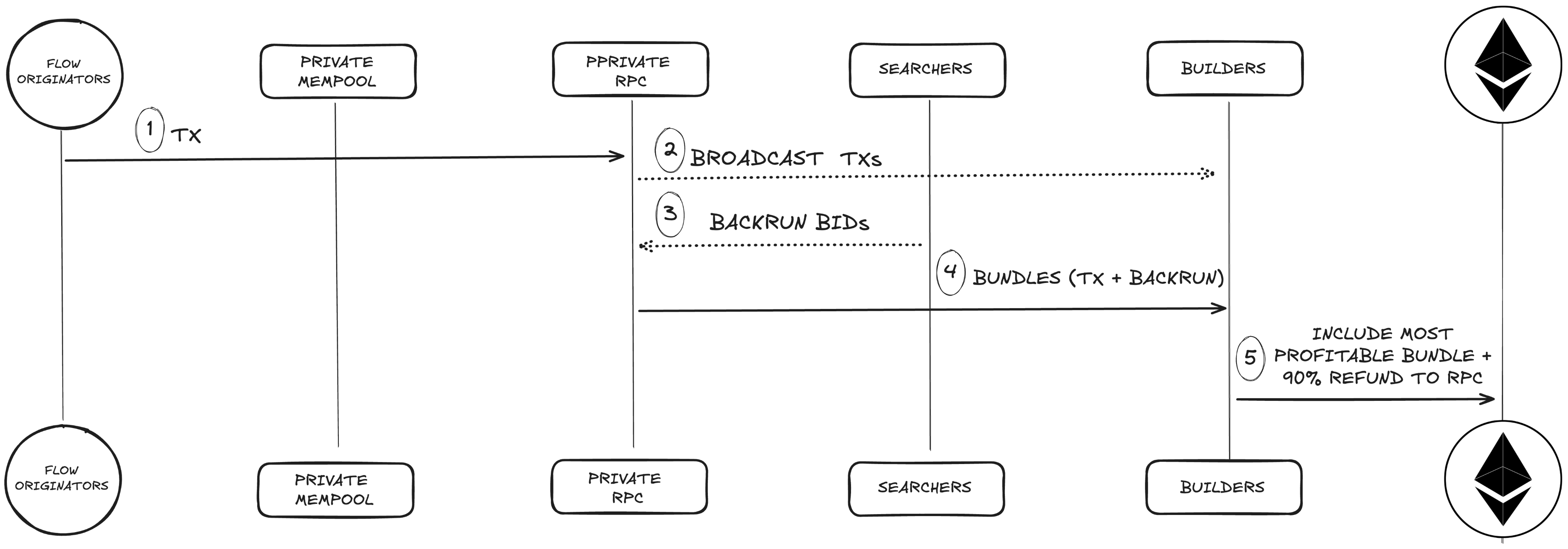}
    \caption{Transaction flow in MEV protection RPCs.}
    \label{fig:ofa_flow}
\end{figure}

As seen in the picture above, Builders connect to a private mempool to source transactions. However, before being shared with builders, the OFA share some partial information about these transactions with searchers. Searchers use the information provided to create  backruns bundles containing a proposed payment (their bid). Because the majority of the bid (90\%) is returned to the user, searchers retain the remaining portion (10\%) of it. This retained amount can be used as a payment to builders—effectively serving as a bribe or incentive—for including their bundle in the block. Private orderflow is an edge builders have from each other, which is why they are all incentivized to connect to these OFAs and play by their rules as if they lose access to such valuable orderflow, it might mean they lost the block creation competition.

There are 4 main RPC providers currently available within Ethereum, MEV Blocker, Flashbots, Blink, and Merkle, who are competing for giving Ethereum's DeFi order flow the best execution and monetization strategy. In order to achieve this, each MEV Protection RPC chose a certain technical infrastructure and auction design for achieving MEV Backruns, as well as a different fee mechanism to monetize their flow. Lets unpack the differences: 

\subsection{Flashbots Protect RPC}
Flashbots Protect is a permissionless RPC launched in 2022. Users can submit transactions without an API key. The RPC selectively shares hints of transaction data with searchers, and the user can configure builder preferences. Flashbots’ builder receives private order flow first, giving it an edge in block building.

\begin{figure}[H]
    \centering
    \includegraphics[width=0.85\textwidth]{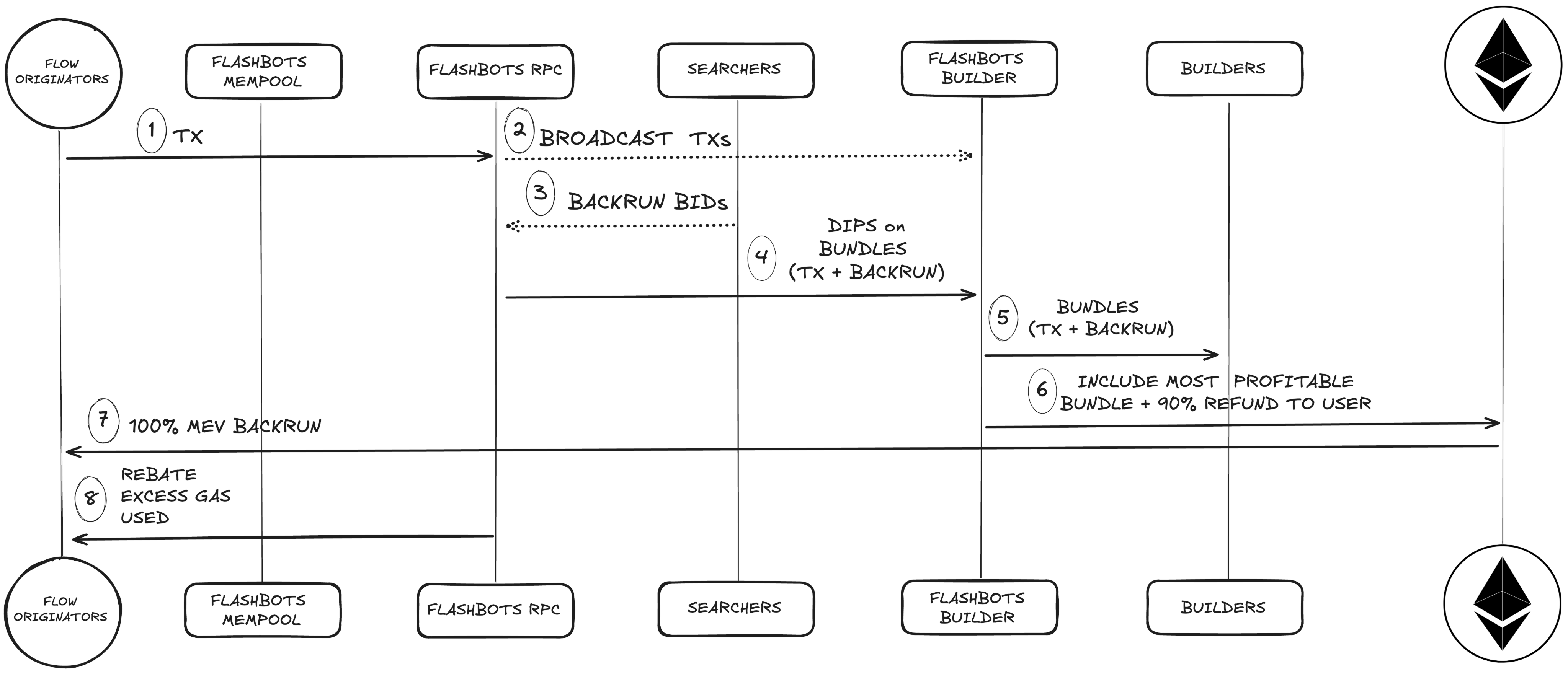}
    \caption{Transaction flow in MEV protection RPCs.}
    \label{fig:ofa_flow}
\end{figure}

\begin{enumerate}
  \item Order flow originators can submit transactions via public endpoints without an API key or third-party permission.
  \item Flashbots RPC shares the transaction with a permission-less set of searchers. The Flashbots MEV-Share node, is responsible for receiving transactions and bundles from users and selectively sharing the transactions "hints" with searchers. 
  \item Transactions are shared with searchers for backrunning. However, the amount of information shared with them as well as Builder selection is not up to the RPC or the searcher but rather the user, which will affect the transaction.
  \item Searchers submit a bundle to Flashbots RPC, who then forwards the transactions first to flashbots builder to give their builder an advantage of private orderflow over the rest of the other builders and aim to win the block.
  \item If the Flashbots builder cant win, then they forward the transactions and bundles to the other connected builders.
  \item Builders receive individual transactions and bundles from Flashbots RPC. They also execute a transaction that pays 90\% of the backrun value to the user. 
  \item Builders execute the user transaction, backruns, and MEV refund, which is automatically sent to tx.origin or the designated address. This ensures users and order flow providers receive 100\% of the refund, with Flasbots taking no cut.
  \item Additionally, Flashbots provides gas rebates to order flow providers by redistributing excess gas fees that are attributable to the orderflow originators every once in a while. However, it is important to note that the calculation on how much “excess gas fee” you will be rebated is not known upfront.
\end{enumerate}

\subsection{Merkle and Blink RPCs}
Launched in late 2023, Merkle and Blink are private RPC services designed to prevent MEV attacks by implementing a permissioned Order Flow Auction (OFA). In both models, only approved searchers can access the private mempool and backrun user transactions. This gated structure enables them to share complete transaction data with trusted searchers and distribute resulting backrun bundles to builders for inclusion.

At a high level, Merkle and Blink follow a similar transaction flow:

\begin{figure}[H]
    \centering
    \includegraphics[width=0.85\textwidth]{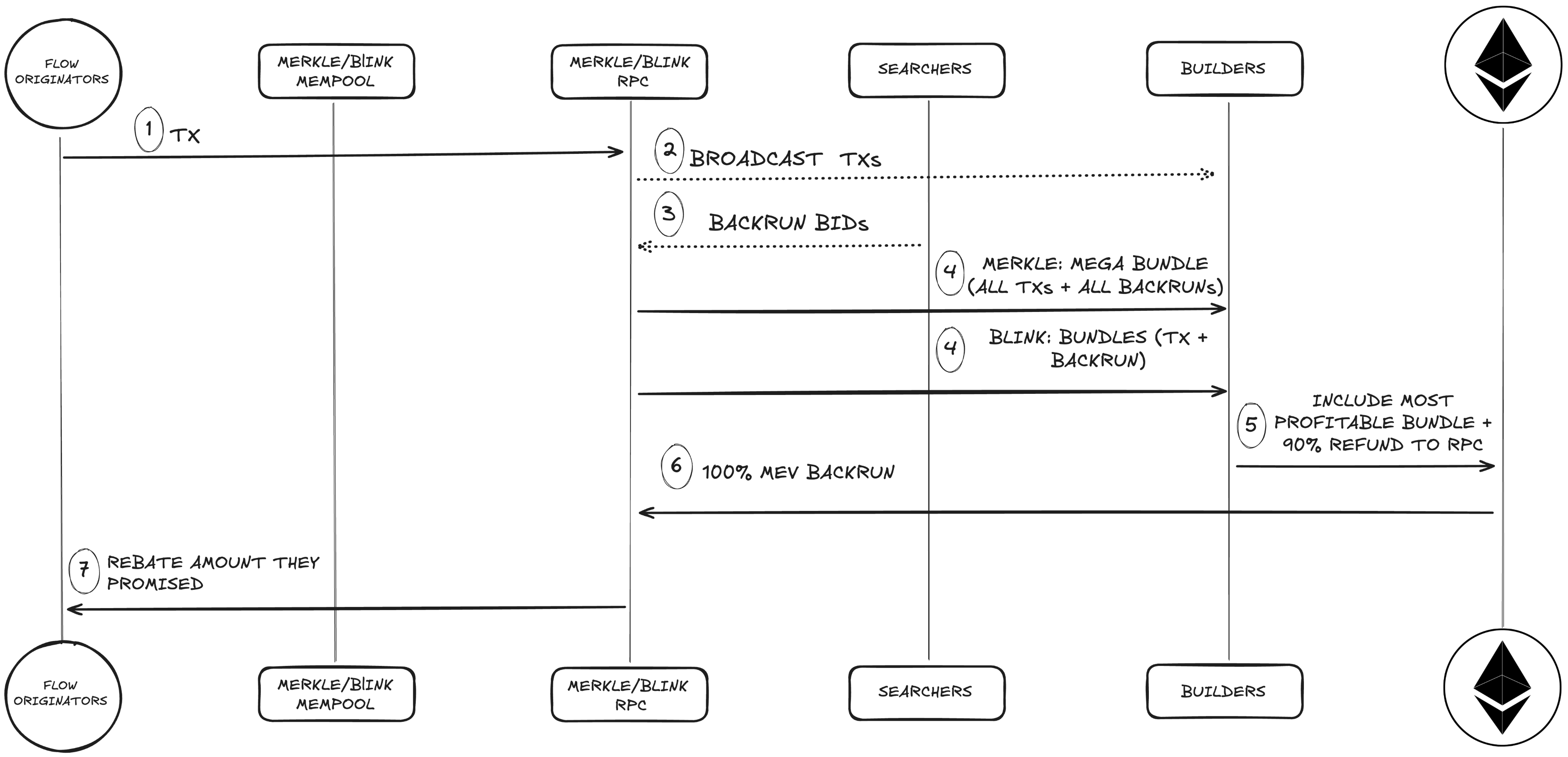}
    \caption{Transaction flow in MEV protection RPCs.}
    \label{fig:ofa_flow}
\end{figure}

\begin{enumerate}
  \item Order flow originators can submit transactions through public endpoints without an API key. However, an API key is required to receive MEV refunds—otherwise, both Merkle and Blink retain the full value extracted.
  \item Transactions are shared with pre-approved searchers who are selected under undisclosed criteria. This creates an opaque environment where access is tightly controlled.
  \item When a backrun opportunity is detected, searchers submit bundles to the RPC. Merkle and Blink select the most profitable bundle to forward, removing builder discretion and increasing the risk of reversion
  \item This is the primary point of divergence:
  - Merkle forwards a single “mega bundle” containing the selected backrun and transaction(s).
  - Blink forwards individual bundles.
  \item Builders are required to include a transaction that pays 90\% of the original transaction's priority fee plus the backrun value to the RPC service. Both RPCs retain MEV profits and manage redistribution. Their per-transaction fee model effectively reduces the transaction’s priority fee from the builder’s perspective, which can slow inclusion and reduce rebate incentives
  \item Builders execute the transactions and pay the MEV backrun to the RPC. Unlike other OFA models, order flow providers do not receive 100\% of the backrun value. Instead, Merkle and Blink pay them in USD and retain any surplus beyond the agreed refund.
  \item Both services offer gas rebates to order flow providers by redistributing part of the builder-paid subscription fees. However, due to the underlying fee structure, receiving a rebate often implies delayed inclusion.
\end{enumerate}

\subsection{MEV Blocker RPC}
Launched in 2023, MEV Blocker is a private RPC designed to prevent MEV attacks by leveraging a permissionless Order Flow Auction (OFA) where any searcher can connect to capture MEV backruns for users transactions that are pending on their private mempool. Because of its permissionless nature, MEV Blocker mixes real transactions with fake transactions so that searchers can receive all the transaction information to perform a backrun and create a bundle, but they can’t take advantage of the information shared as they do not know which transaction will ultimately land onchain. Based on the system design, MEV Blocker RPC transaction flow is a bit different from the other RPCs, and the reason why it has the best connected searchers. Let's examine their transaction flow:

\begin{figure}[H]
    \centering
    \includegraphics[width=0.85\textwidth]{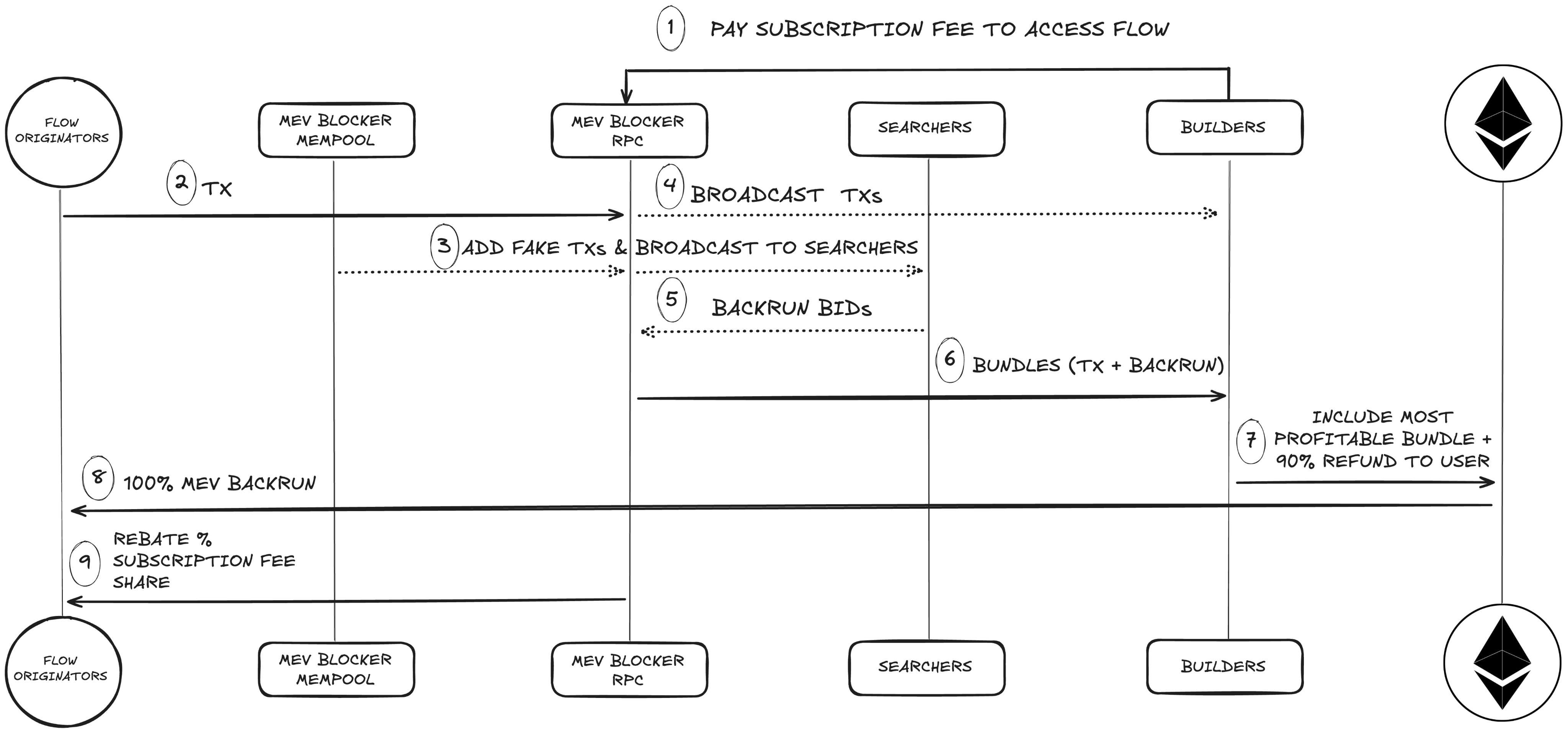}
    \caption{Transaction flow in MEV protection RPCs.}
    \label{fig:ofa_flow}
\end{figure}

\begin{enumerate}
  \item All connected builders pay a weekly fee for MEV Blocker access, calculated by multiplying the number of blocks won by a builder times a ``per-block won fee'' which is common across all builders. The ``per-block won'' fee is calculated using historical data as a percentage of the previous month’s average per-block value of MEV Blocker transactions. The fee is computed monthly and posted onchain~\cite{mevBlocker_smartContract} weekly, and is based on verifiable public data (e.g., Dune query~\cite{mevBlocker_fee_query}).
  \item Order flow originators can submit transactions via public endpoints without an API key or third-party permission.
  \item MEV Blocker RPC shares the transaction (without signature) with a permissioned or permission-less set of searchers (depending on the endpoint). It enhances security by mixing real and AI-generated fake transactions. Since searchers can’t distinguish real from fake, they risk acting on transactions that may never reach the chain, discouraging frontrunning attempts.
  \item Transactions are shared with searchers for backrunning and sent directly to builders for the fastest inclusion.
  \item Searchers are competing for the most profitable option and if they spot a backrun opportunity (real or fake), they submit a bundle to MEV Blocker RPC.
  \item MEV Blocker discards bundles with fake transactions, forwarding only real ones. It also re-adds the transaction signature, preventing searchers from bypassing the system and ensuring MEV rebates reach users. For each transaction included in a bundle, the builder selects the bundle with the highest rebate to the user for on-chain inclusion.
  \item Builders receive individual transactions and bundles from MEV Blocker. Per OFA rules~\cite{OFA_Rules}, they must replace any bundle with a higher-paying one during block-building. They also execute a transaction that pays 90\% of the backrun value to the user. Unlike per-transaction fee models, MEV Blocker’s per-block fee lets builders prioritize inclusion speed and backrun generation without transaction-level cost concerns.
  \item Builders execute the user transaction, backruns, and MEV refund, which is automatically sent to \texttt{tx.origin} or the designated address. This ensures users and order flow providers receive 100\% of the refund, with MEV Blocker taking no cut.
  \item MEV Blocker provides gas rebates to order flow providers by redistributing a percentage of builder subscription fees that are attributable to the orderflow originator.
\end{enumerate}

\section{Methodology}
To measure the effectiveness of each private RPCs, we consider a set of transactions and submit each of them simultaneously multiple times  via each RPC provider.
The goal is to measure the 2 main components of a performant private RPC:

\begin{itemize}
  \item \textbf{Inclusion Quality:} The inclusion quality, that we can split in 2 different metrics: the success rate, i.e. the percentage of transactions submitted which end up on the blockchain, and the time to inclusion, i.e. how long it takes between transaction submission and inclusion of the transaction in a block. This time can be measured in seconds or more preferably in blocks, as this is the real observation. Indeed, Because of the discrete nature of the blocks, submitting 1 second earlier might not make any difference to the time to inclusion, because the next block will not be available any earlier.
  \item \textbf{Execution Quality:} The execution quality. For some transactions, especially swaps, the position where it is included in a block, or the block it is included in may have an impact on the price of the transaction. Therefore, for similar swaps (e.g. sell x tokens A for as many tokens B) across RPC services we measure the execution price to compare their performance. Additionally, searchers can find backrun opportunities, so we also measure the value (measured in ETH) extracted by these backruns to conclude on the quality of each OFAs searcher competition.
\end{itemize}
The study started on block 21344521 (06/12/2024) and ended on block 21637136 (16/01/2025). 
To handle missing values (which emerge when one of the transactions sent via one of the RPC reverts), when measuring price improvement and backrun value we chose one RPC (MEV Blocker) as a reference and compared every other one to this benchmark. Thus we can keep all the transactions that were successfully executed by MEV blocker and at least another RPC, 
Our test includes various types of transactions:
\begin{itemize}
  \item Approvals (13 txs), transfers (139 txs) and supply (1 tx) to AAVE to measure only the inclusion quality.
  \item Various swaps on DEXes, liquid with the pair ETH-USDC (43 txs) or less liquid pairs (77 txs), to also demonstrate the execution quality, i.e. the price execution within the slippage allowance.
\end{itemize}
This period sees various transaction spikes, so some execution failures might be attributed to wrong gas pricing if the transaction takes a couple of blocks to execute.

\begin{figure}[H]
    \centering
    \includegraphics[width=0.85\textwidth]{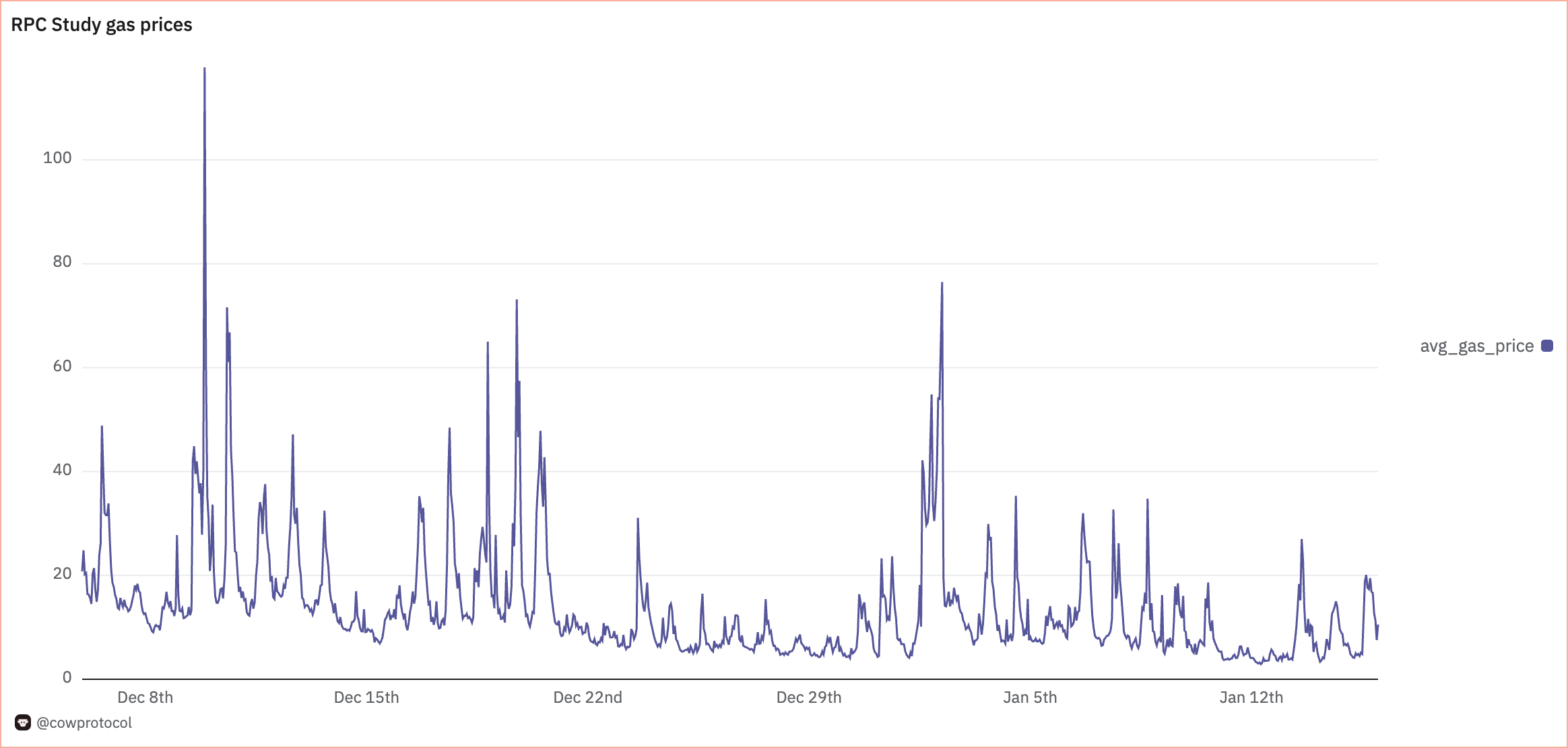}
    \caption{Transaction flow in MEV protection RPCs.~\cite{Gas_Price}}
    \label{fig:ofa_flow}
\end{figure}

\section{Results}
\subsection{Success Rate}
\includegraphics[width=0.9\textwidth]{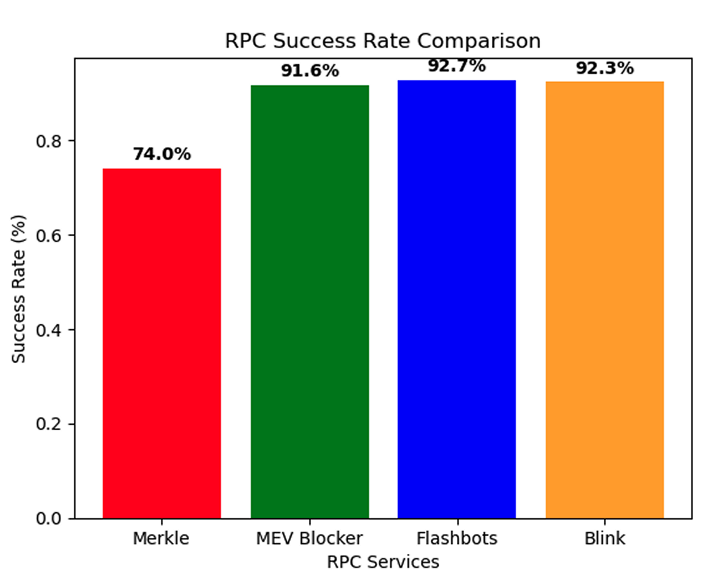}

We define the success rate as the number of transactions that were successfully executed on-chain divided by the number of transactions submitted through an RPC. This implies that all transactions which landed on-chain but reverted are considered as unsuccessful.

Out of the 273 transactions submitted, MEV Blocker, Blink and Flashbots successfully executed their transactions with a rate of exceeding 90\%, while Merkel’s success rate was lower at 74
We assign one color to each RPC that we’ll keep for the following plots.

\subsection{Time to Inclusion}
\includegraphics[width=0.9\textwidth]{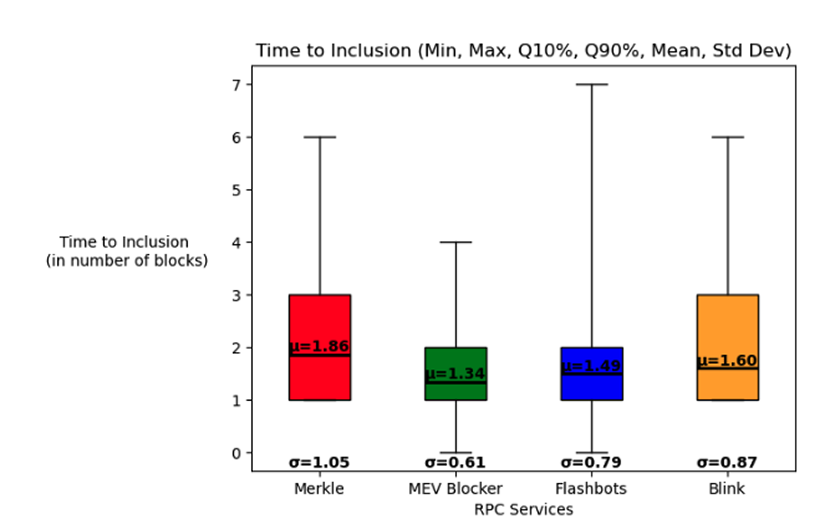}

We now consider the number of blocks between submission and inclusion for the transactions which were successfully executed on-chain. Here, we consider all the successful transactions for each RPC provider, as defined for the success rate.

We plot the box plot of the distribution to time to inclusion (measured in number of blocks), with the minimum, the 10th percentile, the average , the 90th percentile and the maximum value.  represents the standard deviation of the each set. 

We notice that the mean and standard deviation rank the RPC service providers in the same order. This transaction sample demonstrates MEV Blocker fastest execution, followed by Flashbots, then Blink and finally Merkle. 

Flashbot’s execution shows a single large outlier with a time to inclusion of 7 blocks, where the second largest is 4 blocks. This value does not significantly affect the mean and standard deviation, as the sample size is significant enough to compensate for it. With this detail accounted for, the box plot shows consistent performance results.

\subsection{Swap Price Improvement}
Then, we focus only on the swaps and start by observing the quality of the execution which can be measured by the relative price of the swap between the different executions. Here, we compare each RPC service to MEV Blocker, when both have executed the swap. For every trade, we compute price improvement = price(RPC)/price(MEV Blocker) -1, and we measure it in bps. This can be rewritten price improvement = output(RPC)/output(MEV Blocker) -1, with output the amount of tokens received from the swap. Both formulas are similar because the inputs of the transactions are identical. 
The results show that on average Mev Blocker performs better than all other RPCs: it is on average 9bps better than Blink and Merkle, and 21 bps better than  Flashbots. 

We also check the statistical significance of our results by performing a t-test with “H0: price improvement = 0”. Intuitively, the p-value computed by this test represents the probability that, in truth, there is no difference between MEV blocker and the comparison RPC, and therefore the difference we measure in our data is purely random noise. This probability is below 10\% for Flashbot, meaning that the difference between MEV blocker and Flashbot is statistically significant at the 10\% level. The difference between MEV blocker and the other 2 RPCs instead is not significant at any standard significance level.

\begin{figure}[H]
    \centering
    \includegraphics[width=0.85\textwidth]{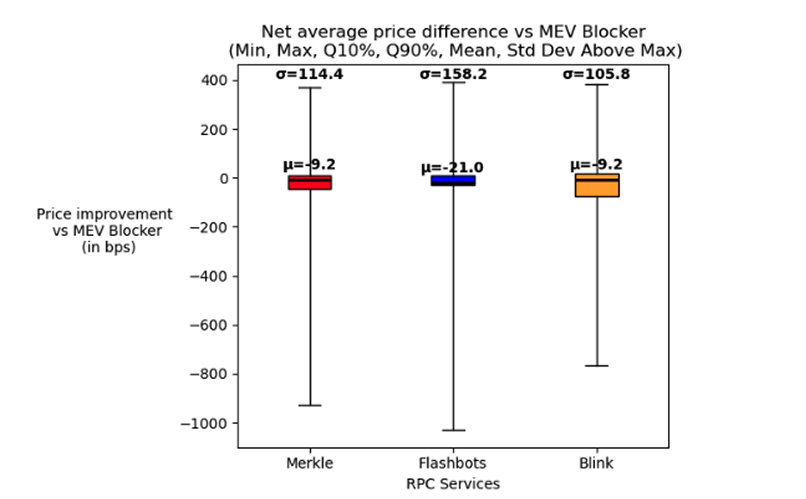}
    \caption{Transaction flow in MEV protection RPCs.}
    \label{fig:ofa_flow}
\end{figure}

\begin{table}[h]
\centering
\begin{tabular}{lcc}
\toprule
Sample Size & t-test (ETH) & p-value \\
\midrule
Merkle 92 & -0.764 & 0.223 \\
Flashbots 98 & -1.309 & 0.097 \\
Blink 97 & -0.851 & 0.198 \\
\bottomrule
\end{tabular}
\caption{Net Average Price Difference vs MEV Blocker.}
\end{table}

\subsection{Backrun quality}

First, we measure the backruns generated by the RPCs:

\begin{table}[h]
\centering
\begin{tabular}{lcccc}
\toprule
RPC Provider & MEV Blocker & Merkle & Flashbots & Blink \\
\midrule
Backrun (in ETH) & 0.00350 & 0.00108 & 0.00088 & 0.00173 \\
\bottomrule
\end{tabular}
\caption{Backrun value per RPC.}
\end{table}

We follow a similar methodology as for swap price improvements: we compare between each RPC solution and MEV Blocker, and only consider swaps  that landed on-chain both via MEV blocker and via at least another RPC.

We plot a similar box plot as previously. The first plot represents the average backrun, and its distribution for all the transactions. The second plot only includes transactions where one of them (MEV Blocker or the RPC) has a non-0 backrun.

The swaps can be classified in 2 categories: very liquid pairs (ETH-USDC) and less liquid pairs (ETH-GNO, ETH-COW, and ETH-AAVE). The liquid pairs did not generate any backruns, which leads to many 0 values. Indeed, for these pairs there is enough liquidity so that one trade of reasonable size does not unbalance the pool enough to create an arbitrage opportunity. Additionally, even the less liquid pairs don’t generate backruns every time. Therefore, we plot the values with and without the transactions with no backrun. For this analysis, we find that MEV Blocker’s backruns are larger than any other RPCs solutions with more than 10\% confidence.

However, if we look at the backruns received by the users, the values are different as Blink and Merkle don’t send these backruns back to the user:

\begin{table}[h]
\centering
\begin{tabular}{lcccc}
\toprule
RPC Provider & MEV Blocker & Merkle & Flashbots & Blink \\
\midrule
Backrun (in ETH) & 0.00350 & 0 & 0.00088 & 0 \\
\bottomrule
\end{tabular}
\caption{User Rebate per RPC.}
\end{table}

The comparison between MEV Blocker and Flashbots remains unchanged, because both RPCs send 100\% of the rebates to their users. However, we can conclude that MEV Blocker and Flashbots are better for the users than Merkl and Blink, in terms of rebates value.

\begin{figure}[H]
    \centering
    \includegraphics[width=0.85\textwidth]{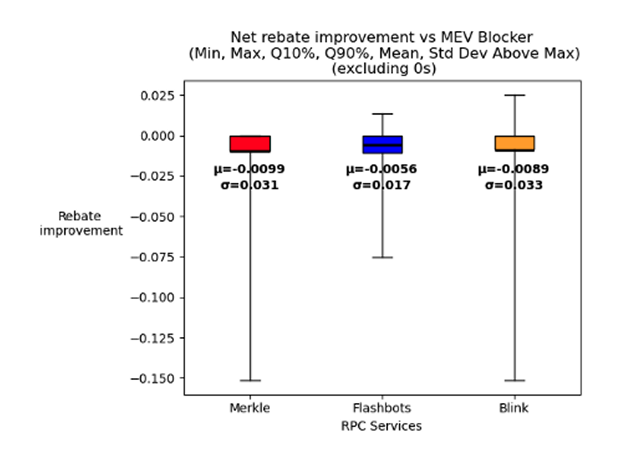}
    \caption{Transaction flow in MEV protection RPCs.}
    \label{fig:ofa_flow}
\end{figure}

\begin{figure}[H]
    \centering
    \includegraphics[width=0.85\textwidth]{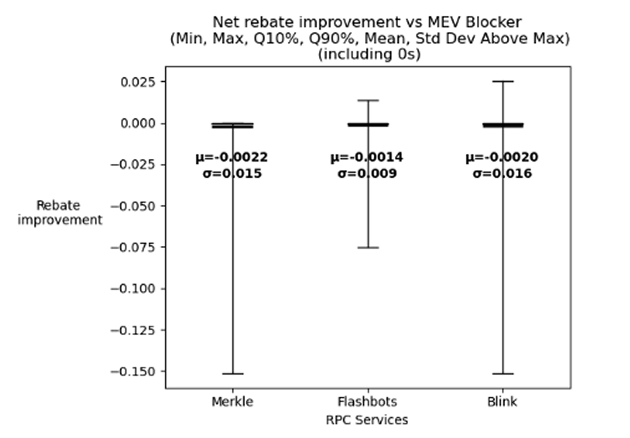}
    \caption{Transaction flow in MEV protection RPCs.}
    \label{fig:ofa_flow}
\end{figure}

\begin{table}[h]
\centering
\begin{tabular}{l|ccc|ccc}
\toprule
 & \multicolumn{3}{c|}{Including 0} & \multicolumn{3}{c}{Excluding 0} \\
\textbf{RPC} & \textbf{Sample size} & \textbf{T-test} & \textbf{P-value} & \textbf{Sample size} & \textbf{T-test} & \textbf{P-value} \\
\midrule
Merkle   & 120 & -1.586 & 5.765e-2 & 27 & -1.624 & 5.826e-2 \\
Flashbots & 120 & -1.761 & 4.042e-2 & 30 & -1.811 & 4.085e-2 \\
Blink    & 120 & -1.373 & 8.623e-2 & 27 & -1.391 & 8.800e-2 \\
\bottomrule
\end{tabular}
\caption{T-test and p-value comparisons including and excluding zero backrun values}
\label{tab:t_test_comparison}
\end{table}

\section{Conclusion}
From these comparisons, we can rank each RPC service provider based on the different metrics. MEV Blocker stands out for being the only RPC that is either the best or among the best in all metrics. . Flashbot ranks similarly or just below MEV blocker in  time to inclusion and backrun value. However, its execution price is the lowest among all RPCs. Finally Merkle and Blink both generate very low rebates. Also, Merkle stands out for its  low success rate.

\begin{table}[h]
\centering
\begin{tabular}{lcccc}
\toprule
RPC & Inclusion Delay & Price Improvement & Average Rebate & Success Rate \\
\midrule
MEV Blocker & 1.34 blocks & -- & 0.0035 ETH & 85\% \\
Flashbots & 1.49 blocks & 21bps less & 0.0016 ETH & 84\% \\
Merkle & 1.86 blocks & 9bps less & 0.0009 ETH & 70\% \\
Blink & 1.56 blocks & 9bps less & 0.0000 ETH & 84\% \\
\bottomrule
\end{tabular}
\caption{Net Average Price Difference vs MEV Blocker.}
\end{table}

\newpage
\section*{Appendix}
\subsection{Explanation Between the Different Fee Models}
This section compares how the transaction flow of some OFAs are prioritized by builders in the event that builders pay out money to the RPCs:

\begin{itemize}
  \item \textbf{Flashbots RPC Mempool} – Free access; Flashbots builders get a first look before transactions are shared with other builders. Transactions are prioritized by gas fees.
  \item \textbf{Merkle RPC Private Mempool} – Per-transaction fee (90\% priority tax); transactions are shared with all MEV Boost builders who deduct fees from transaction value.
  \item \textbf{Blink RPC Private Mempool} – Per-transaction fee (90\% priority tax); transactions are shared with all MEV Boost builders who deduct fees from transaction value.
  \item \textbf{MEV Blocker Private Mempool} – Fixed monthly fee; transactions are valued as-is and shared with all MEV Boost builders, with fees deducted from proposer overall MEV received.
\end{itemize}

\begin{figure}[H]
    \centering
    \includegraphics[width=0.85\textwidth]{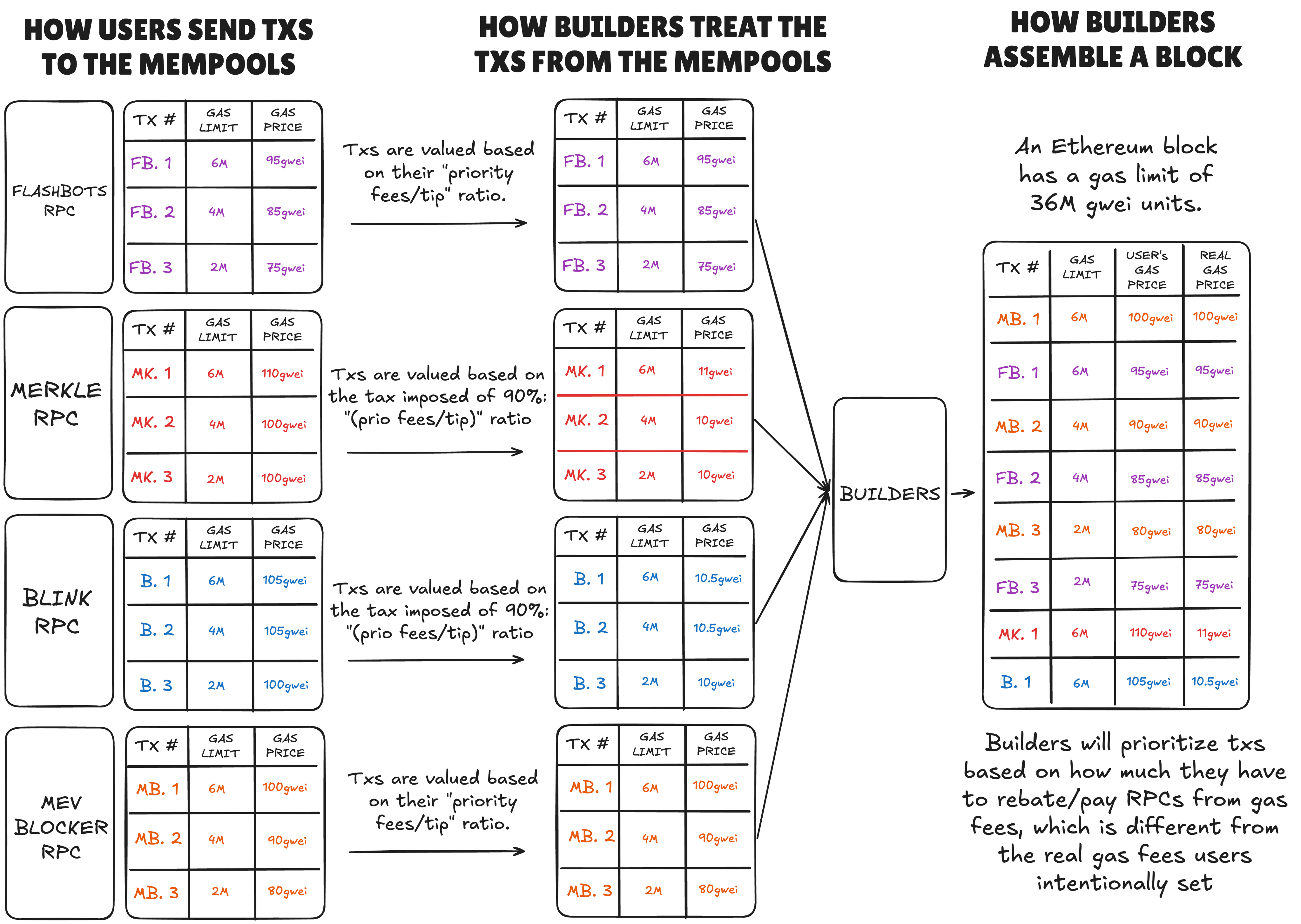}
    \caption{Transaction flow in MEV protection RPCs.}
    \label{fig:ofa_flow}
\end{figure}

A key factor in block construction is Ethereum’s 36M gas limit. If total gas usage exceeds this, builders must prioritize transactions based on their economic value, which varies by mempool access model and monetization strategy. 

\subsection*{Builder Transaction Example}
Builders’ available transactions:

\begin{itemize}
  \item \textbf{Flashbots RPC}: (1) 6M gas at 95 gwei; (2) 4M gas at 85 gwei; (3) 2M gas at 75 gwei
  \item \textbf{Merkle RPC}: (1) 6M gas at 110 gwei; (2) 4M gas at 100 gwei; (3) 2M gas at 100 gwei
  \item \textbf{Blink RPC}: (1) 6M gas at 105 gwei; (2) 4M gas at 105 gwei; (3) 2M gas at 100 gwei
  \item \textbf{MEV Blocker RPC}: (1) 6M gas at 100 gwei; (2) 4M gas at 90 gwei; (3) 2M gas at 80 gwei
\end{itemize}

\textbf{Selected Transactions for Inclusion:}
\begin{itemize}
  \item MEV Blocker: (1) 6M at 100 gwei; (2) 4M at 90 gwei; (3) 2M at 80 gwei
  \item Flashbots: (1) 6M at 95 gwei; (2) 4M at 85 gwei; (3) 2M at 75 gwei
  \item Merkle: (1) 6M at 110 gwei
  \item Blink: (1) 6M at 105 gwei
  \item \textbf{Not Included} –  Merkle: (2) 4M at 100 gwei; (3) 2M at 100 gwei (excluded due to per-transaction fees)
  \item \textbf{Not Included} – Blink: (2) 4M at 105 gwei; (3) 2M at 100 gwei (excluded due to per-transaction fees)
\end{itemize}
As you can see in the example, Builders ultimately prioritize in terms of priority fee paid (caveat that all transactions are executing the same action), which means that monetizing the RPC flow via one type of OFA vs.\ another will actually have an impact on how transactions land, and when. In this case, you can see that $\frac{2}{3}$ of both Merkle and Blink RPC transactions were excluded due to per-transaction fees, reducing profitability for the Order Flow provider. Builders favored MEV Blocker and Flashbots transactions, which incur no extra costs to them (because Flashbots has no monetization strategy for its flow, builders can include transactions as they see fit). This highlights a key trade-off: MEV Blocker’s subscription model maximizes inclusion, while Merkle’s and Blink’s per-transaction fees deter selection in constrained block space. Not only that, but they also lower the position in the block in which they will include the transaction, because, after all, Builders prioritize transactions with the highest net value after fees, making some RPC monetization models less competitive.

Builders prioritize transactions by net value after fees. The example shows 2/3 of Merkle and Blink’s transactions were excluded due to lower profitability. Flashbots and MEV Blocker flows were favored due to fewer additional costs. This illustrates how OFA monetization strategies impact execution timing and inclusion reliability.

\subsection*{Flashbots Model Builders Treatment}
Flashbots RPC does not have a fee mechanism in place to monetize orderflow, and therefore builders do not need to take anything into account when treating their txs.

\begin{figure}[H]
    \centering
    \includegraphics[width=0.85\textwidth]{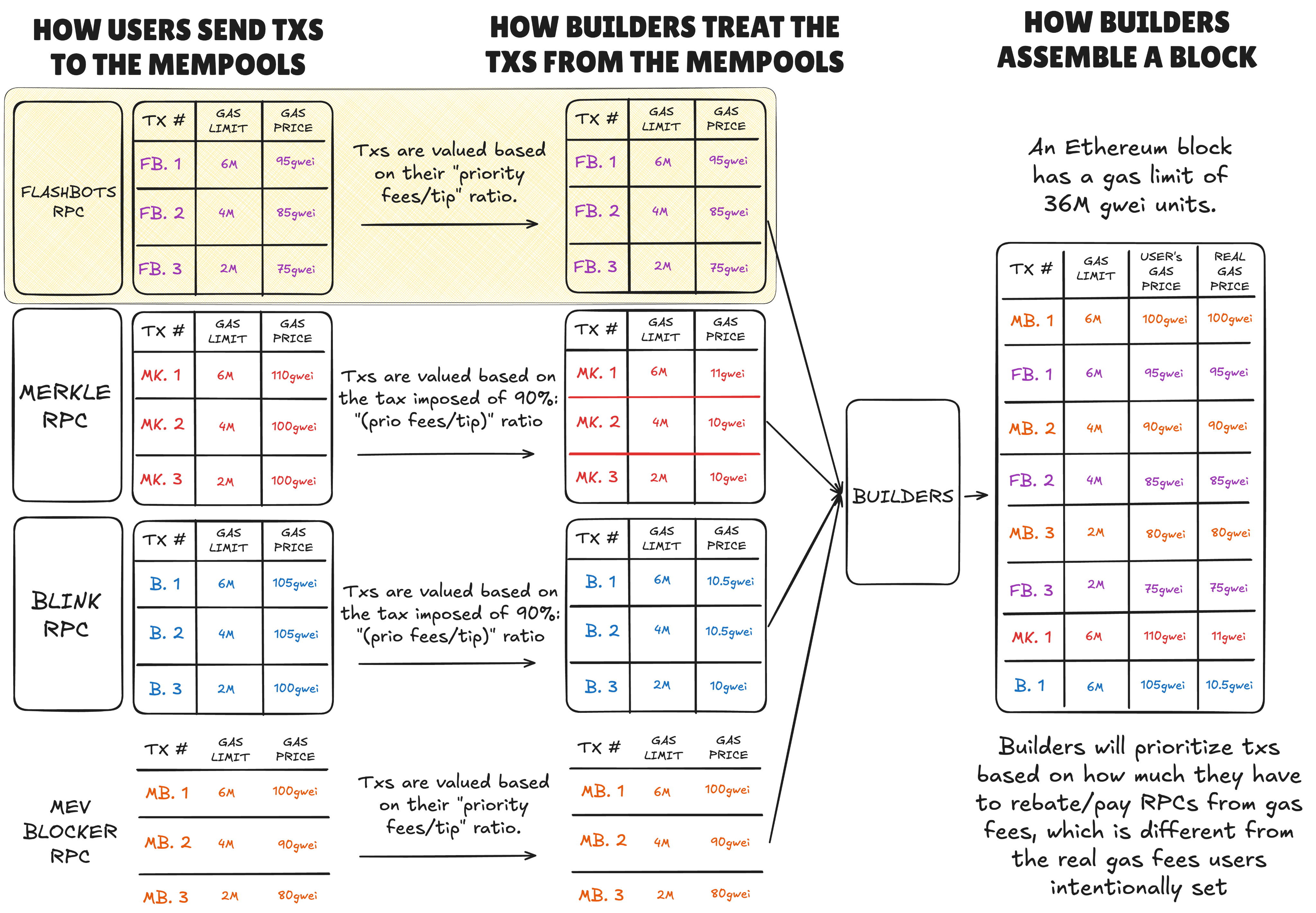}
    \caption{Transaction flow in MEV protection RPCs.}
    \label{fig:ofa_flow}
\end{figure}

However, its important to note two aspects of Flashbots models that differentiate from other OFA. 
\begin{itemize}
  \item Builder selection: the user has the right to select the amount of builders it wants its transactions shared with, which means that transactions might not be included as fast as possible…
  \item Flashbots Builder gets 1st look at the transactions. Before sharing with all other builders, Flashbots builders get priority at trying to do something with the private orderflow from their RPC, if their builder cant improve the situation, then the flow is shared with the rest of the builders. This means that transactions will be slightly delayed, as by the time other builders see the transactions after Flashbot's builder has “rejected” them, they might be irrelevant.
\end{itemize}

\subsection*{Merkle’s Fee Model Builders Treatment}
This section explains how builders treat the fee mechanism implied by Merkles RPC OFA.

\begin{figure}[H]
    \centering
    \includegraphics[width=0.85\textwidth]{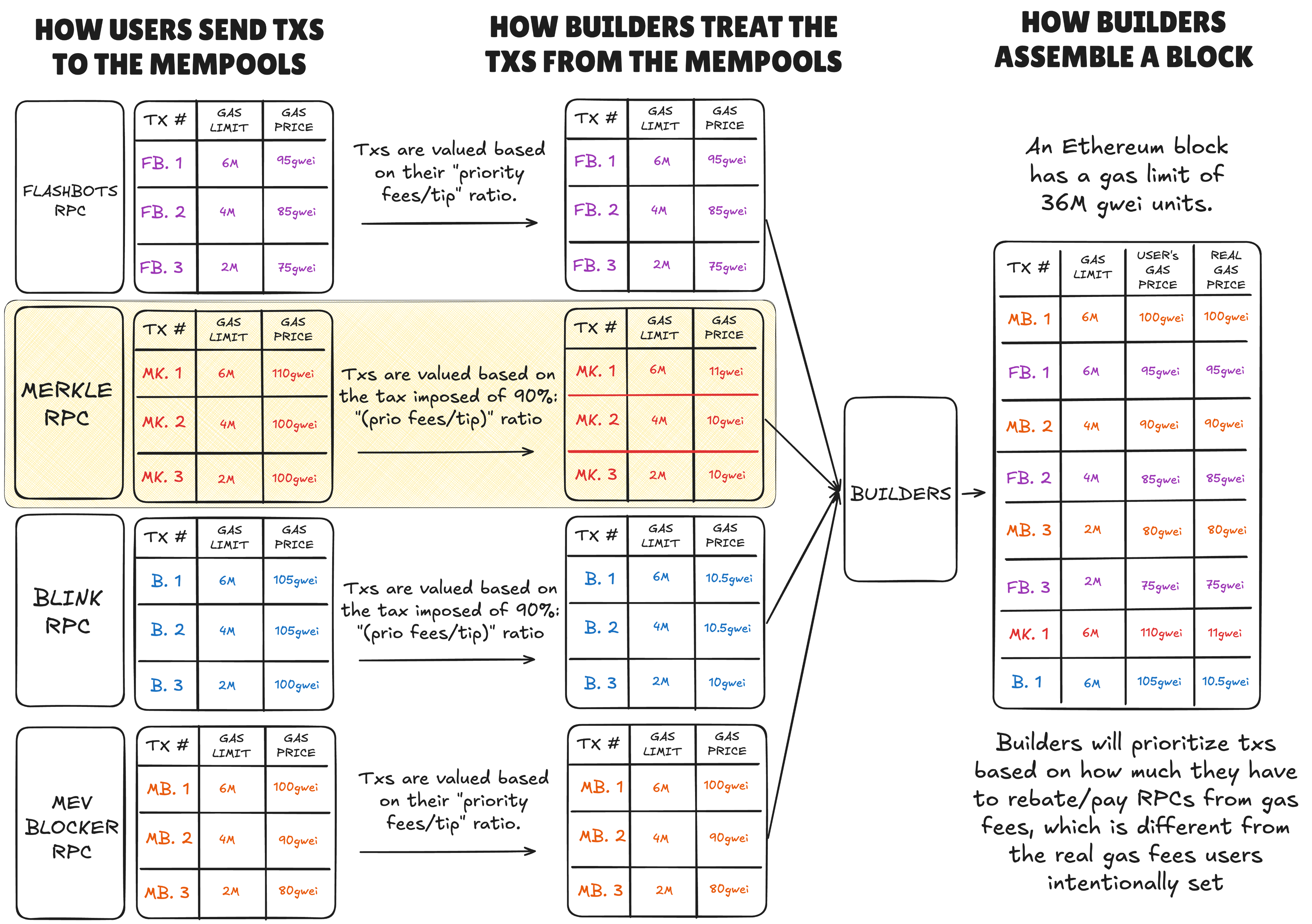}
    \caption{Transaction flow in MEV protection RPCs.}
    \label{fig:ofa_flow}
\end{figure}

Builders connected to Merkle pay a per-transaction fee:
\begin{equation*}
  \text{Fee} = (\text{priority fee}) \times (1 - \text{refundPercent})
\end{equation*}

Why this fee mechanism? This fee model allows Merkle to extract all the value, both MEV and priority fees from order flow providers users, and then decide how much they want to kick back to them. However, as shown in the example, and in our tests, this fee mechanism indeed forces Builders to delay transactions, as they need to pay the RPC a certain amount of money that comes from the transaction itself.

\subsection*{Blink’s Fee Model Builders Treatment}
This section explains how builders treat the fee mechanism implied by Blink RPC OFA.

\begin{figure}[H]
    \centering
    \includegraphics[width=0.85\textwidth]{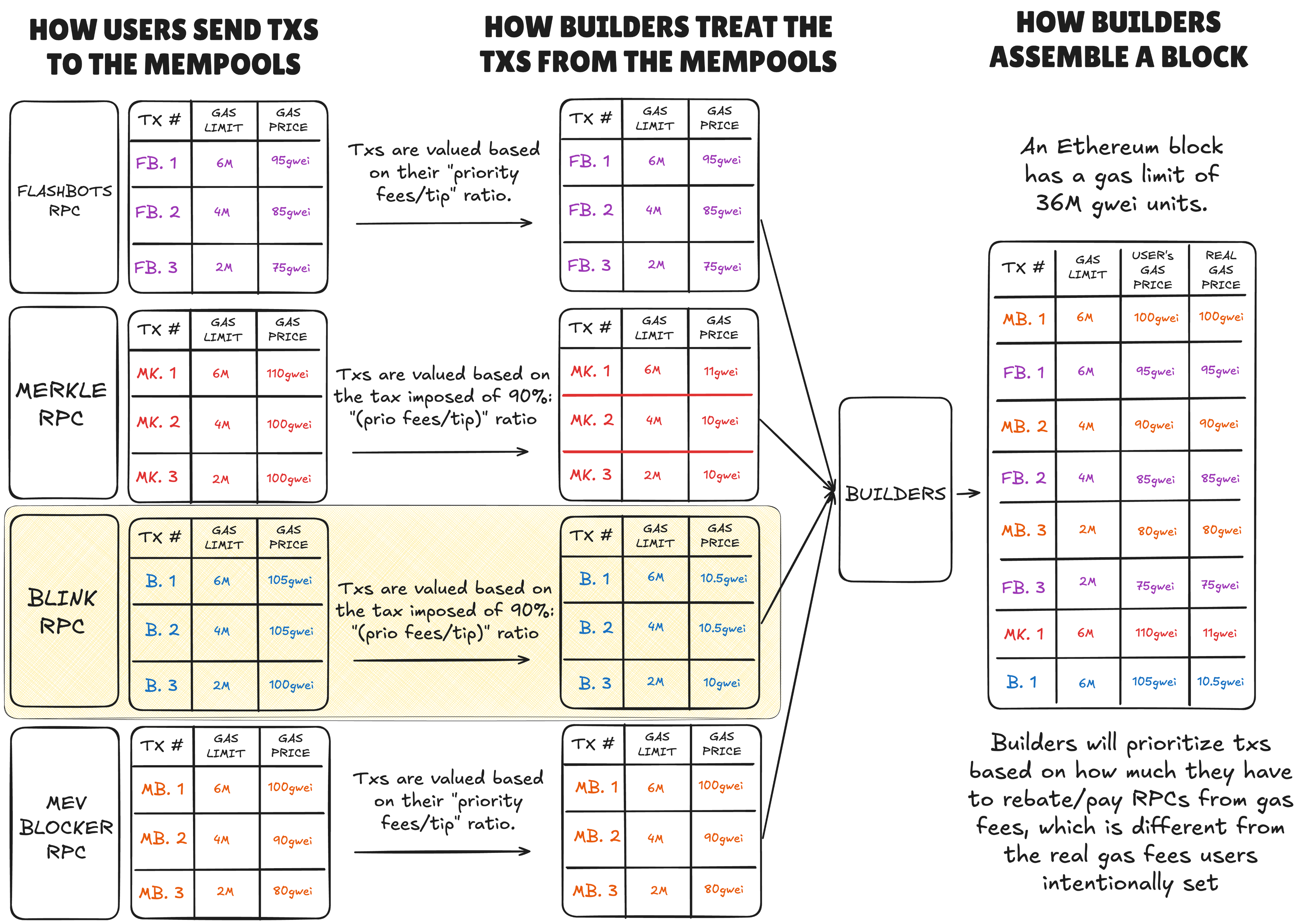}
    \caption{Transaction flow in MEV protection RPCs.}
    \label{fig:ofa_flow}
\end{figure}

Same as Merkle. Builders pay:
\begin{equation*}
  \text{Fee} = (\text{priority fee}) \times (1 - \text{refundPercent})
\end{equation*}

Why this fee mechanism? This fee model allows Blink to extract all the value, both MEV and priority fees from order flow providers users, and then decide how much they want to kick back to them. However, as shown in the example, and in our tests, this fee mechanism indeed forces Builders to delay transactions, as they need to pay the RPC a certain amount of money that comes from the transaction itself.

\subsection*{MEV Blocker Fee Model Builders Treatment}
This section explains how builders treat the fee mechanism implied by MEV Blocker OFA.

\begin{figure}[H]
    \centering
    \includegraphics[width=0.85\textwidth]{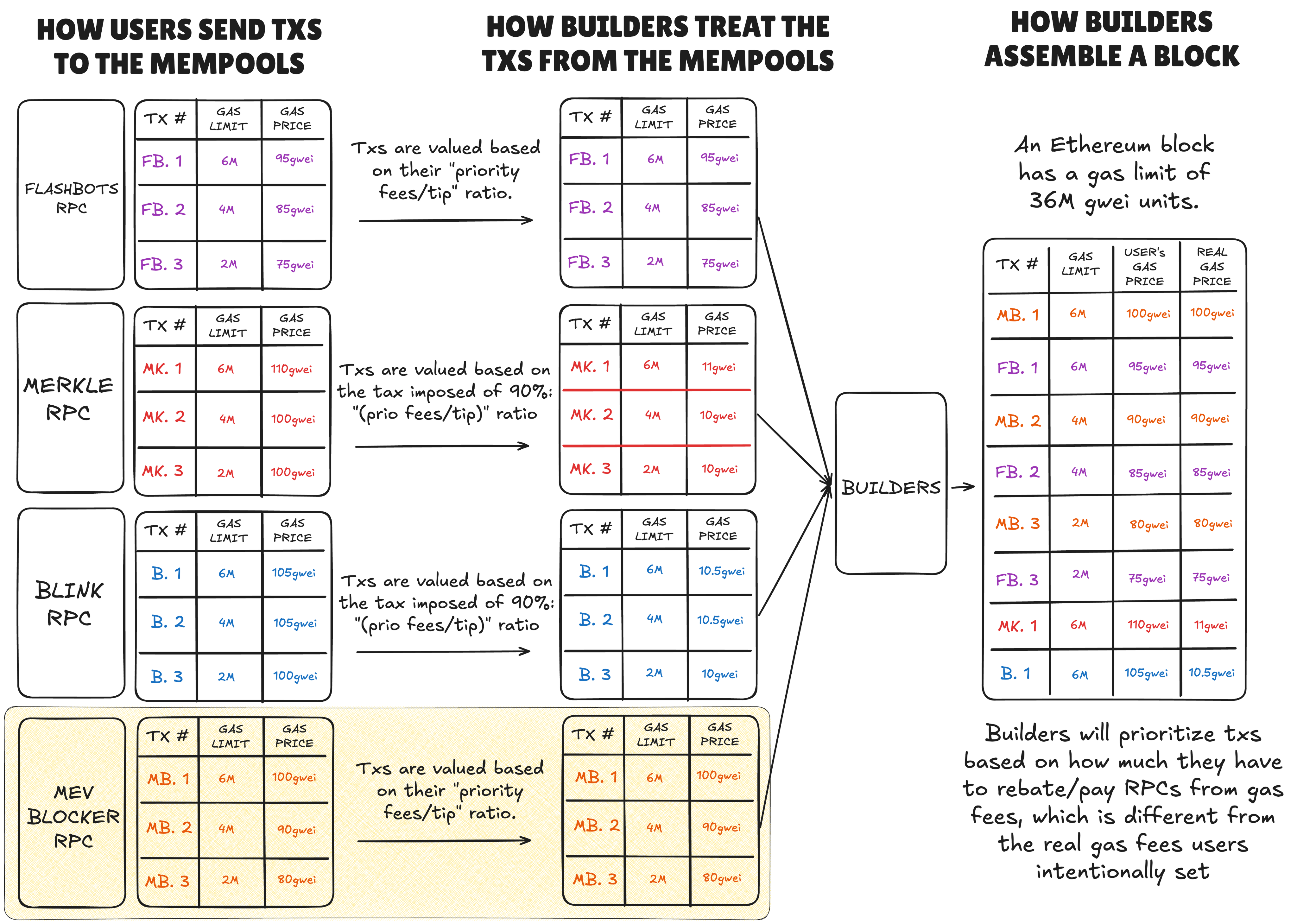}
    \caption{Transaction flow in MEV protection RPCs.}
    \label{fig:ofa_flow}
\end{figure}

MEV Blocker uses a monthly per-block-won fee:
\begin{equation}
    M_{t-1} = \frac{A_{t-1} - B_{t-1}}{C_{t-1}}
\end{equation}

Where

\begin{itemize}
    \item $A_{t-1}$ = Total MEV Blocker transaction value during period $t - 1$, excluding rebates.
    \item $B_{t-1}$ = Total value of MEV Blocker transactions also in the mempool during period $t - 1$.
    \item $C_{t-1}$ = Number of blocks mined by builders receiving MEV Blocker transactions during period $t - 1$.
\end{itemize}

Builders using MEV Blocker pay a per-block-won fee, recalculated monthly as a percentage of the previous period’s MEV Blocker transaction flow (priority fees + backruns) value. This fee is independent of which transactions they include or their order. Why this fee mechanism? Because it's designed to be non-distortionary, meaning it does not affect which transactions a builder includes. A builder winning a block always pays the same fee, regardless of which MEV Blocker transactions are used or their order. The model is also accurate over time—if builders remain connected and win a similar share of blocks, they pay, with a slight delay, the average MEV transaction value from the previous period. Lastly, this mechanism reduces payments from builders to validators, improving efficiency while maintaining fairness.
If a builder is charged a fee $f$ per block won, winning becomes worth $f$ less, so they lower their max bid by the same amount. However, since all MEV Blocker connected builders adjust equally, rankings remain unchanged—the highest bidder before the fee remains the highest after. Winning a block isn’t just about collecting priority fees; it also helps builders maintain market share, so they factor in this benefit when setting their bids (known as a max subsidy).
Since the auction is ascending, the winner only needs to outbid the second-highest bidder slightly. Because all bids drop by $f$ the final winning bid is also $f$ lower. The winning builder then pays $f$ to MEV Blocker, meaning their total cost remains unchanged. Builders do not need to adjust transaction selection or inclusion speed—the fee does not impact their

\end{document}